\documentclass[epjST]{svjour}
\usepackage{graphicx}
\usepackage{epstopdf}  
\usepackage{amsmath}
\usepackage{amssymb}
\usepackage{enumitem}   
\newcommand{\be}{\begin{equation}}
\newcommand{\ee}{\end{equation}}
\newcommand{\x}{{\bf x}}
\newcommand{\Mf}{{\bf M_f}}
\newcommand{\Msb}{{\bf M_{St}^{b}}}
\newcommand{\Mfb}{{\bf M_{f}^{b}}}
\newcommand{\Mst}{{\bf M_{St}}}  
\newcommand{\Msynt}{{\bf M_{synt}}}
\newcommand{\Mtau}{{\bf M}_\tau}
\newcommand{\C}{\textsf{\textbf{C}}}
\newcommand{\diag}{\mbox{diag}}
\newcommand{\dd}{\boldsymbol\delta}
\newcommand{\A}{\textsf{\textbf{A}}}
\newcommand{\xii}{\boldsymbol\xi}
\newcommand{\R}{\textsf{\textbf{R}}}
\begin{document}
\title{A map for heavy inertial particles in fluid flows}
\author{Rafael D. Vilela\inst{1}\fnmsep\thanks{\email{rafael.vilela@ufabc.edu.br}} \and Vitor M. de Oliveira\inst{2} }
\institute{CMCC, Universidade Federal do ABC, Santo Andr\'e, SP, Brazil \and Instituto de F\'\i sica, Universidade de S\~ao Paulo, S\~ao Paulo, SP, Brazil}
\abstract{
We introduce a map which reproduces qualitatively many fundamental properties of the dynamics of heavy particles in fluid flows. These include a uniform rate of decrease of volume in phase space, a slow-manifold effective dynamics when the single parameter $s$ (analogous of the Stokes number) approaches zero, the possibility of fold caustics in the ``velocity field'', and a minimum, as a function of $s$, of the Lyapunov (Kaplan-Yorke) dimension of the attractor where particles accumulate.   
} 
\maketitle
\section{Introduction}
\label{intro}
The transport of inertial (i.e., finite-size) particles in fluid flows is ubiquitous in nature and important for the industry \cite{springer}. Especially important is the case of particles denser than the carrying flow, called aerosols or heavy particles. Raindrops in clouds and dust particles in gas nebulae are examples of heavy particles in fluid flows, and the formation of rain and planets are processes intimately related to their dynamics.

The general equation of motion for a finite-size particle is known since the work of Maxey and Riley \cite{maxeyriley}. It can be interpreted as Newton's law with the various terms corresponding to Stokes drag, buoyancy, and other forces. The dynamics takes place in a phase space of dimension $2N$, where $N$ is the dimension of the fluid flow, which reflects the inertial character of the particles: at a given instant, the velocity is not a function of position for a finite-size particle. In the case of heavy particles, drag is usually the dominant force and the resulting dynamics displays a number of interesting features as the control parameter, called the Stokes number, $St$ (essentially a measure of the inertia of the particle), is varied.  First, for small enough $St$, the effective phase space is a $N$-dimensional slow manifold  and a synthetic approximation for the dynamics can be derived \cite{maxey87,haller&sapsis,haller&sapsis2}. Second, particles distribute inhomogeneously in physical space due to different effects, which is usually referred to as {\em preferential concentration} \cite{maxey87}. This effect is most pronounced for particles with $St$ of the order of unity \cite{bec,wilkclustering}. Third, for $St$ larger than a threshold of the order of unity, caustics can be formed in phase space, leading to multivalued velocity fields and a pronounced increase in the collisional rate of the particles \cite{sling,wilkcaustics}. 

Nonlinear Dynamics has a long tradition in replacing differential equations by maps (i.e., difference equations) in the modeling of natural phenomena. This is no less the case for transport in fluid flows. Sommerer \cite{sommerer} used random maps to model the behavior of floaters in chaotic flows which was experimentally addressed in \cite{ottscience}. Cartwright {\it et al.} \cite{cartwright1} defined a discrete-time dynamics called {\em bailout embedding} and used it to investigate the qualitative behavior of neutrally buoyant particles in chaotically advected flows \cite{cartwright2} (see also \cite{gupte}). Among the advantages of using maps, simplicity and clarity stand out. Besides, maps are more suitable for a prompt numerical investigation which can indicate promising directions to pursue with the integration of the differential equations they mimic. 

In this paper, we introduce a map, which we call the {\em Stokes map}, that mimics the equation of motion for a heavy particle in a fluid flow. We show that the above-mentioned and other important features of the true (i.e., continuous-time) dynamics are reproduced by the Stokes map.

\section{The Stokes map}
\label{sec:1}
In the absence of gravity, the equation of motion for a small spherical particle much denser than the background fluid reads \cite{maxeyriley}:
\begin{equation}
\dot{{\bf v}}=\frac{1}{St}[\bf{u}(\bf{x},t)-\bf{v}], 
\label{maxey}
\end{equation}
where ${\bf v}$ is the velocity of the particle, $\bf{u}(\bf{x},t)$ is the fluid velocity field evaluated at the particle position $\bf{x}\in\mathbb{R}^{\it N}$ and time $t$, and $St$ is the Stokes number of the particle. The latter can be expressed as $St=(2a^2U)/(9\nu L R)$, where $a$ is the particle's radius, $\nu$ is the kinematic viscosity of the fluid, $U$ and $L$ are the characteristic velocity and length of the flow, respectively, and $R=\frac{2\rho_f}{\rho_f +2\rho_p}$ depends on the densities of particle ($\rho_p$) and fluid ($\rho_f$). It is worth noting that Eq.~(\ref{maxey}) is a simplification which neglects the history force. As shown in \cite{daitche2014}, the history force is often important even for very heavy particles. In the case of water droplets in air ($R\approx 10^{-3}$), Eq.~(\ref{maxey}) is an accurate (within a 2$\%$ margin) advection equation for aerosols provided that $St < 0.1$ \cite{daitche2014}.

Equation (\ref{maxey}) together with $\dot{{\bf x}}={\bf v}$ defines a $2N$-dimensional flow, the divergence of which gives the uniform rate of decrease of phase space volumes: $-N/St$.  The case $St=0$ corresponds to a singular limit which recovers the $N$-dimensional dynamics of fluid tracers, while the limit $St\to \infty$ corresponds to free-particle dynamics. We have built a simple map which reproduces the discrete-time versions of these features. The Stokes map is given by:
\begin{equation}
\left\{
\begin{array}{ccl}
\x_{n+1}=\x_n + \dd_{n},     \cr

\dd_{n+1}=s\hspace{1mm}\dd_n + (1-s)[\Mf(\x_{n+1}) - \x_{n+1}] .
\end{array}
\right. 
\label{mapa} 
\end{equation}
The position vector at the $n$-th iterate is given by $\x_n$, while $\dd_n$ is the discrete-time counterpart of the particle velocity (integrated over one iterate). We shall adopt an abuse of terminology from now on and refer to $\dd_n$ also as the particle velocity. Both $\x_n$ and $\dd_n$ are $N$-dimensional vectors, as is the map $\Mf$ which gives the dynamics of fluid tracers.
The parameter $s$ is the analogous of the Stokes number and its range is the interval $[0,1)$. As is the case of $St$ in Eq.~(\ref{maxey}), $s$ is conceived to be a monotonically increasing function $s(a)$ of the particle's radius with $s(0)=0$. Note, however, that $s \to 1$ as $a\to \infty$ (compare with the corresponding limit $St\to\infty$). Since the aim of the Stokes map is to provide a {\em qualitative} discrete-time proxy for the true  dynamics, the details of $s(a)$ are unimportant. We now describe the qualitative similarities between the dynamics given by Eq.~(\ref{maxey}) and Eq.~(\ref{mapa}).

\begin{enumerate}[label=\roman*)]
\setlength\itemsep{1em}
\item {\em Clear limit cases of fluid tracers and free particles.---}  If $s=0$, Eq.~(\ref{mapa}) reduces to the $N$-dimensional map $\x_{n+1}=\Mf(\x_n )$, which is the dynamics for fluid tracers. This is so because the second line in Eq. (2) implies that $\dd_n = \Mf (\x_{n}) - \x_{n}$ for  $s=0$. On the other hand, the limit $s\to1$ leads to the $2N$-dimensional map $(\x_{n+1},\dd_{n+1})=(\x_n + \dd, \dd)$ which describes the dynamics of a free particle of constant velocity $\dd$. We note in passing that, for $0<s<1$, the vector $\dd_{n+1}$ is linearly interpolated between the values it assumes in the aforementioned limit cases.

\item {\em Uniform decrease of phase space volumes.---} 
In the case of difference equations, the rate of change of phase space volumes is given by the absolute value of the determinant of the Jacobian of the map. If we define $\boldsymbol\xi_n\equiv(\x_n, \dd_n)$, Eq.~(\ref{mapa}) can be rewritten as $\xii_{n+1}=\Mst(\xii_n)$, where $\Mst \in \mathbb{R}^{2{\it N}}$ is the Stokes map. The determinant of its Jacobian can be easily calculated and yields $\det[{\bf D}\Mst(\boldsymbol\xi)]=s^{N}$, which is constant. This implies that the shrinkage of volumes in phase space is uniform, as happens with the continuous-time dynamics.
 
\item {\em Slow-manifold dynamics.---} When $St\ll1$, the phase space of inertial particles obeying Eq.~(\ref{maxey}) rapidly collapses to a $N$-dimensional slow manifold \cite{haller&sapsis} and the effective dynamics is described by a synthetic velocity field \cite{maxey87} given to first order in the Stokes number by 
\begin{equation}
 \dot{{\bf x}}={\bf u}-St\left[\frac{\partial {\bf u}}{\partial t}+{\bf
      u}\cdot\nabla{\bf u}\right].
\label{synthmaxey}      
\end{equation}
In a similar fashion, when $s\ll1$ the phase space of inertial particles obeying Eq.~(\ref{mapa}) rapidly collapses to a $N$-dimensional slow manifold and the effective dynamics is described by a synthetic map which, to first order in $s$, reads:
\be
\x_{n+1}=\Msynt(\x_n ) \equiv \Mf(\x_n )+s\left[2\x_n -\Mf(\x_n )-\Mf^{-1}(\x_n )\right].
\label{synthmap}
\ee
Equation~(\ref{synthmap}) can be derived by noting that the second line of Eq.~(\ref{mapa}) can be rewritten as $\dd_{n}=\Mf(\x_{n}) - \x_{n}+s[\dd_{n-1} -\Mf(\x_{n}) +\x_{n}] $ and that $\dd_{n-1}=\Mf(\x_{n-1}) - \x_{n-1}+\mathcal{O}(s)=\x_{n}-\Mf^{-1}(\x_n )+\mathcal{O}(s)$.

The slow manifold can be fully visualized in the case of one-dimensional fluid dynamics. This is so because the phase space is then two-dimensional. For this reason, let us consider the one-dimensional compressible fluid map given by $x_{n+1}=M_{f} (x_n )\equiv x_{n}^{3}+x_{n}/2$.  Figure \ref{fig:1} shows the slow manifold traced by inertial  particles after a few iterates of the Stokes map as well as its first-order approximation given by Eq.~(\ref{synthmap}). We see that the synthetic first-order approximation is in excellent agreement with the actual manifold for values of $s$ as small as 0.01.

\begin{figure}
\resizebox{0.99\columnwidth}{!}{%
\includegraphics{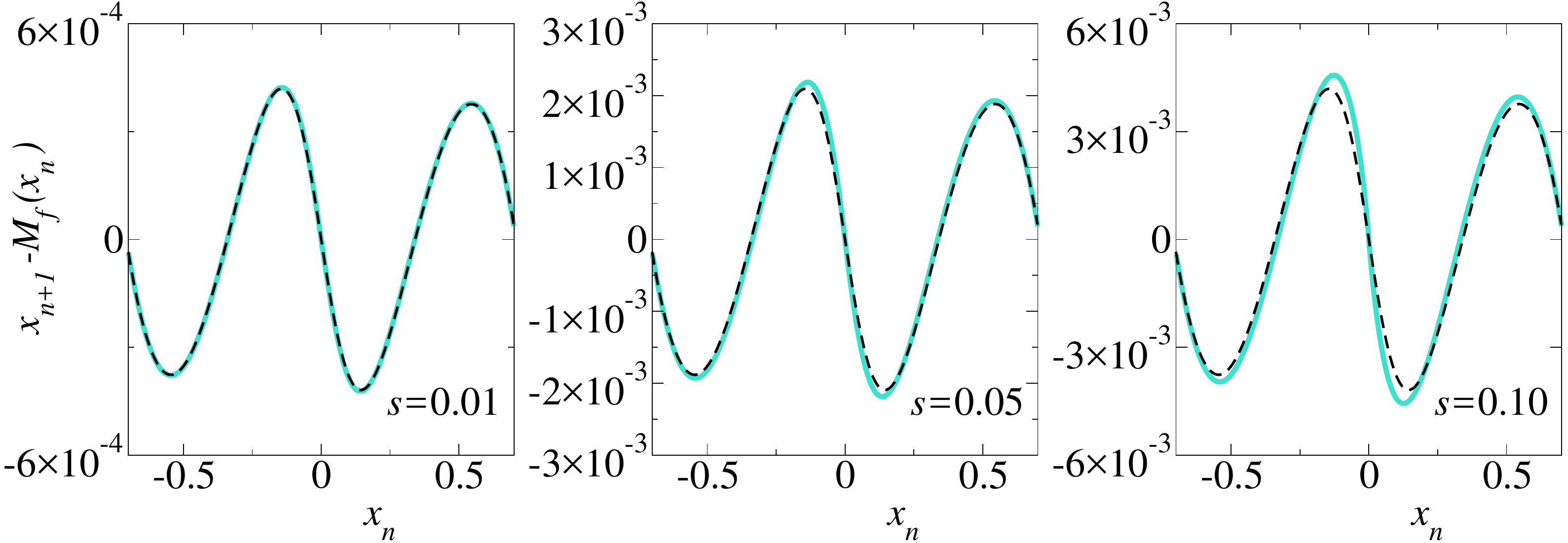} }
\caption{Slow manifold (continuous turquoise line) where heavy particles (initial position $\x_{0} \in(-1/\sqrt{2},1/\sqrt{2})$ and initial velocity equal to fluid velocity, i.e., $\dd_{0}=\Mf(\x_{0})-\x_{0}$) accumulate after a few ($n=5$) iterations of the Stokes map for $M_f (x_n )= x_{n}^{3}+x_{n}/2$. The first-order synthetic approximation of the slow manifold corresponds to the dashed black line. For $s=0.01$ (left panel), the synthetic approximation matches the slow manifold to the line width. For $s=0.05$ (middle panel), mild discrepancies can be noticed in the valleys and peaks and for $s=0.10$ (right panel) the mismatch increases as higher order terms grow.}
\label{fig:1}       
\end{figure}

\item{\em Centrifuge effect.---} 
Equation (\ref{synthmaxey}) has been shown to imply that, in steady or slowly-varying flows, small-$St$ inertial particles tend to exit regions dominated by vorticity and accumulate in regions dominated by strain. This is so because the divergence of the synthetic field is proportional to squared vorticity minus squared strain \cite{maxey87physfluids}. This centrifuge effect is one of the possible mechanisms leading to preferential concentration (two other mechanisms are discussed below). 

The centrifuge effect takes place continuously in time. It is therefore conceptually difficult to be addressed in a discrete-time dynamics. Consider for instance a fluid dynamics given by the identity map in two dimensions. It could equally well model still fluid (a ``flow'' with no motion at all) and a rigid rotation over an angle of $2\pi$. When inertial particles are placed in both flows, the centrifuge effect is of course present only in the second one. Having pointed this out, let us focus on the classes of steady and slowly-varying incompressible two-dimensional flows and restrict ourselves to the case where the fluid map $\Mf$ corresponds to a time-$\tau$ stroboscopic map (i.e., defined by sampling the fluid flow at time instants $t_o$, $t_o +\tau$, $t_o +2\tau$ and so on), to be called $\Mtau$, with $\tau$ smaller than the characteristic time scale of the flow (such as the turnover time of vortical structures). Incompressible two-dimensional flows are formally Hamiltonian systems, implying that $\det[{\bf D}\Mtau]=1$. Two cases are then typical, defined by the character of the eigenvalues.

If the eigenvalues of ${\bf D}\Mtau(\x)$ are real, we have ${\bf D}\Mtau(\x)=\C\A\C^{-1}$, where $\C$ is a conjugacy matrix,  $\A=\diag(\lambda,\lambda^{-1})$ and $\lambda \in \mathbb{R}$ is a function of $\x$. 
In this case, straightforward algebra leads to the eigenvalues of ${\bf D}\Msynt (\x)$, which are given by $(1 - s + 2 \lambda s - \lambda^2 s)/\lambda$ and $(\lambda^2 - s + 2 \lambda s - \lambda^2 s)/\lambda$.
The rate of change of phase space volumes is then given by the absolute value of the following expression:
$$
\det[{\bf D}\Msynt (\x)]=1 - 2 s - \frac{s}{\lambda^2} + \frac{2 s}{\lambda} + 2 \lambda s - \lambda^2 s + 
  6 s^2 + \frac{s^2}{\lambda^2} - \frac{4 s^2}{\lambda} - 4 \lambda s^2 + \lambda^2 s^2.
$$
For concreteness, let $\lambda$ be the largest eigenvalue. It can be shown that $1<\lambda<2$ is a sufficient condition for the absolute value of $\det[{\bf D}\Msynt (\x)]$ to be smaller than 1 for any $s$ in the interval $(0,1)$, implying area contraction in physical space. Note that $s=0$ corresponds to the limit of fluid tracer dynamics and in this case the above determinant is equal to 1, a consequence of area preservation.

In the case of complex conjugate eigenvalues of ${\bf D}\Mtau(\x)$, they must be unitary, implying that ${\bf D}\Mtau(\x)=\C\R\C^{-1}$, where $\C$ is a conjugacy matrix and $\R=\left(\begin{array}{ccl}
\cos\theta && \sin\theta     \cr
-\sin\theta && \cos\theta
\end{array}\right)$ is a rotation matrix. The corresponding eigenvalues of ${\bf D}\Msynt (\x)$ are given by $2s + \cos\theta - 2s \cos\theta + i \sin\theta$ and its complex conjugate. We then have:
$$
\det[{\bf D}\Msynt (\x)]=1 - 2 s + 6 s^2 - 4 s (-1 + 2 s) \cos\theta + 
 2 (-1 + s) s \cos2 \theta.
$$
It can be shown that $\theta \in (-\pi/2,\pi/2)-\{0\}$ is a sufficient condition for the absolute value of $\det[{\bf D}\Msynt (\x)]$ to exceed 1 for any $s\in(0,1)$, implying area dilation in physical space. Here again $s=0$ corresponds to the limit of fluid tracer dynamics which preserves area.

Taken together, the results above imply a net ``centrifuge effect'' for the map $\Msynt$ provided the time interval $\tau$ which defines the stroboscopic fluid map $\Mtau$ is sufficiently small in the following precise terms: it should yield expansion factors smaller than 2 in strain-dominated regions of the flow (from which $\Mtau$ is derived) and rotation angles smaller than $\pi/2$ in vorticity-dominated regions. In this sense should $\tau$ be  ``smaller than the characteristic time scale'' of the underlying flow. 

It is important to point out that the centrifuge effect exists for any positive $St$ in Eq.~(\ref{maxey}), including large $St$ for which Eq.~(\ref{synthmaxey}) does not apply. Interestingly, the Stokes map given by Eq.~(\ref{mapa}) also mimics the centrifuge effect for any $s \in (0,1)$ under the same conditions stated for $\Msynt$ in the previous paragraph, as we now argue. Let $\Mf$ correspond to a rotation by an angle smaller than $\pi/2$ and let $(\x_{0}, \dd_{0})$ correspond to the initial condition of a heavy particle with initial velocity equal to fluid velocity, i.e., $\dd_{0}=\Mf(\x_{0})-\x_{0}$. Then the observation that the vector $\dd_{1}$ is linearly interpolated between the vectors $\dd_{0}$ and $\Mf(\x_{1})-\x_{1}$ and simple geometry lead to the conclusion that the one-iterate jump from $\x_{1}$ to $\x_{2}$ corresponds to the heavy particle moving outwards.

The centrifuge effect is very striking in steady cellular flows, where particles are expelled from the elliptic stagnation points and accumulate in the regions of the separatrices which connect the hyperbolic stagnation points. Figure \ref{fig:2} shows this effect for particles obeying Eq.~(\ref{mapa}) with $\Mf=\Mtau$, where $\tau=0.015$ and the underlying flow corresponds to Hamilton's equations $\dot{x}=\partial \psi/\partial y$ and $\dot{y}=-\partial \psi/\partial x$ for the stream function $\psi=\sin(2\pi x)\sin(2\pi y)$. Note that $s=0.02 \ll 1$ for the particles shown in this figure, which implies that simulations using Eq.~(\ref{mapa}) or Eq.~(\ref{synthmap}) lead to very similar results. 

\begin{figure}
\begin{center}
 \resizebox{0.75\columnwidth}{!}{%
\includegraphics{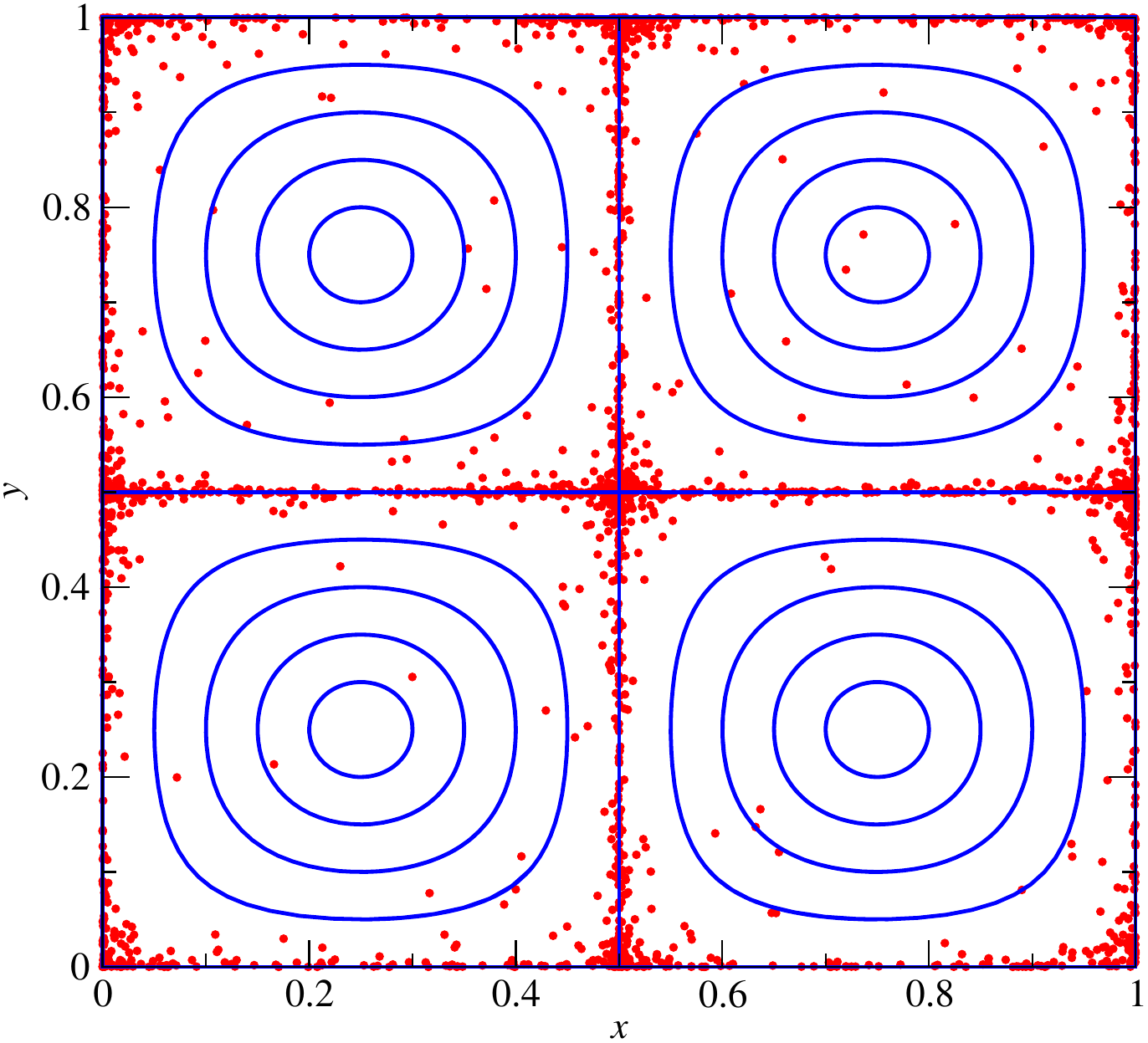} } 
\end{center}
\caption{Centrifuge effect for inertial particles ($s=0.02$) under the Stokes map for $\Mf$ defined as a (numerically obtained) stroboscopic map ($\tau=0.015$) of the steady cellular flow corresponding to the stream function $\psi=\sin(2\pi x)\sin(2\pi y)$. Inertial particles (red dots) accumulate along the separatrices and hyperbolic stagnation points. The number of inertial particles displayed is 2000 and they are initially uniformly distributed in the unit square with velocity equal to the local velocity of the fluid, i.e., $\dd_{0}=\Mf(\x_{0})-\x_{0}$. Streamlines are shown for reference (continuous blue lines).}
\label{fig:2}       
\end{figure}

\item{\em Caustics.---} As discussed above, for small $St$ (and $s$), the dynamics of inertial particles is effectively $N$-dimensional and takes place in a slow manifold corresponding, in the continuous-time case, to the synthetic field in Eq.~(\ref{synthmaxey}) and, in the discrete-time case, to the synthetic map given by Eq.~(\ref{synthmap}). As the parameter $St$ (or $s$) increases, the distances from those first-order approximations to their corresponding actual slow manifolds also increase. That mismatch is initially due to the increasing importance of higher order terms, which could in principle be included in the synthetic approximation. However, for sufficiently large $St$ (or $s$),  {\em fold caustics} \cite{wilkcaustics} develop, which completely derail any expansion in that parameter. This effect is again clearly visible for 1D maps. Figure~\ref{fig:3} shows it for the same map as considered in Fig.~\ref{fig:1}. 

\begin{figure}
\resizebox{0.99\columnwidth}{!}{%
  \includegraphics{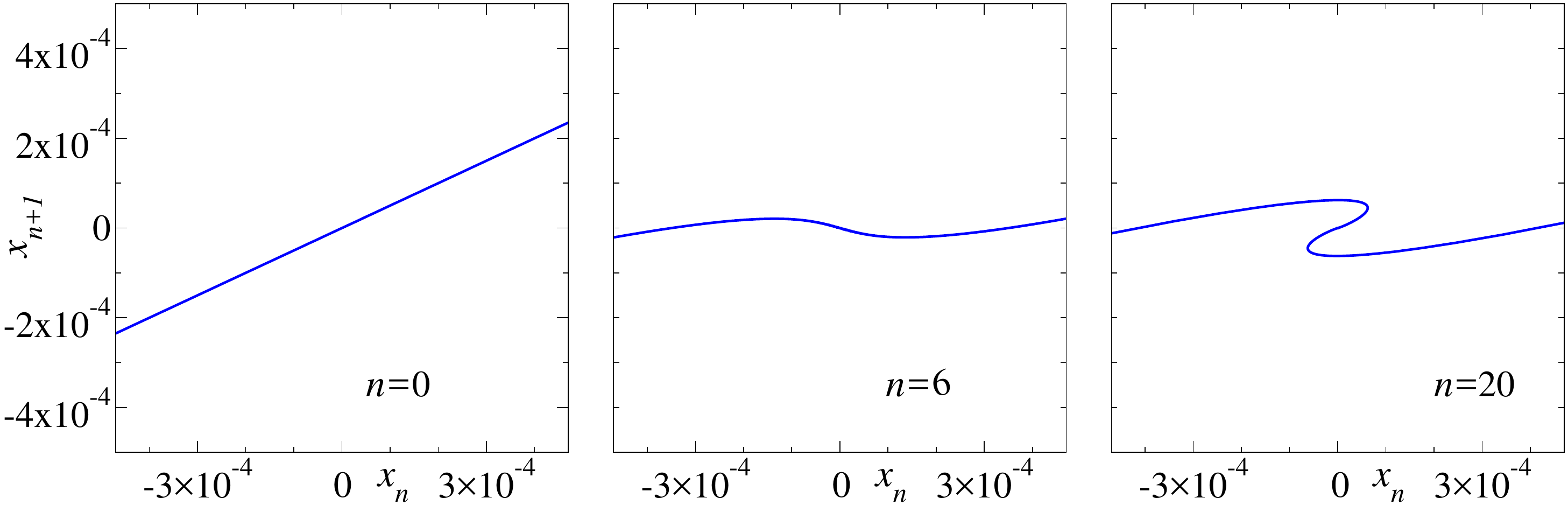} }
\caption{Caustics formation in 1D maps: $x_{n+1}$ as a function of $x_{n}$ for an ensemble of inertial particles ($s=0.15$) under the Stokes map with initial ($n=0$, left panel) velocity equal to the local fluid velocity. Already at $n=6$ (middle panel) particles experience a pronouncedly different dynamics and at $n=20$ (right panel) fold caustics are observed, leading to multivalued velocity fields. The fluid map $M_f$ is the same as the one in Fig.~\ref{fig:1}.}
\label{fig:3}       
\end{figure}

The formation of caustics constitutes another mechanism for preferential concentration. This can be easily understood as follows: assuming a smooth particle density along the slow manifold in phase space, it is primarily the slope of the manifold which causes a density variation in the projection of the manifold onto physical space. Caustics are just the outcome of increasing that slope to the point of folding. The effect is dramatic at the folding points, where the density of particles diverges \cite{wilkcaustics}. 

Besides preferential concentration, caustics provide another cause for the increase of the rate of collisions of inertial particles: multivalued velocity fields, which means that particles at the same position can have different velocities. In one dimension, the discrete-time version of this effect can be seen on the right panel of Fig.~\ref{fig:3}. Multivalued velocity maps also occur in higher dimension. Figure~\ref{fig:4} illustrates this phenomenon in the case where the fluid dynamics is given by the blinking vortex-source system \cite{arefblink,telkarolyi}. This is a 2D ideal flow corresponding to two alternating steady vortex-sources located at different positions. A desired feature displayed by this flow is that the equations of motion for fluid tracers can be analytically integrated to yield a map for the fluid tracers (see derivation in \cite{telkarolyi}). Its expression is given in Appendix A. 

\begin{figure}
\resizebox{0.99\columnwidth}{!}{%
  \includegraphics{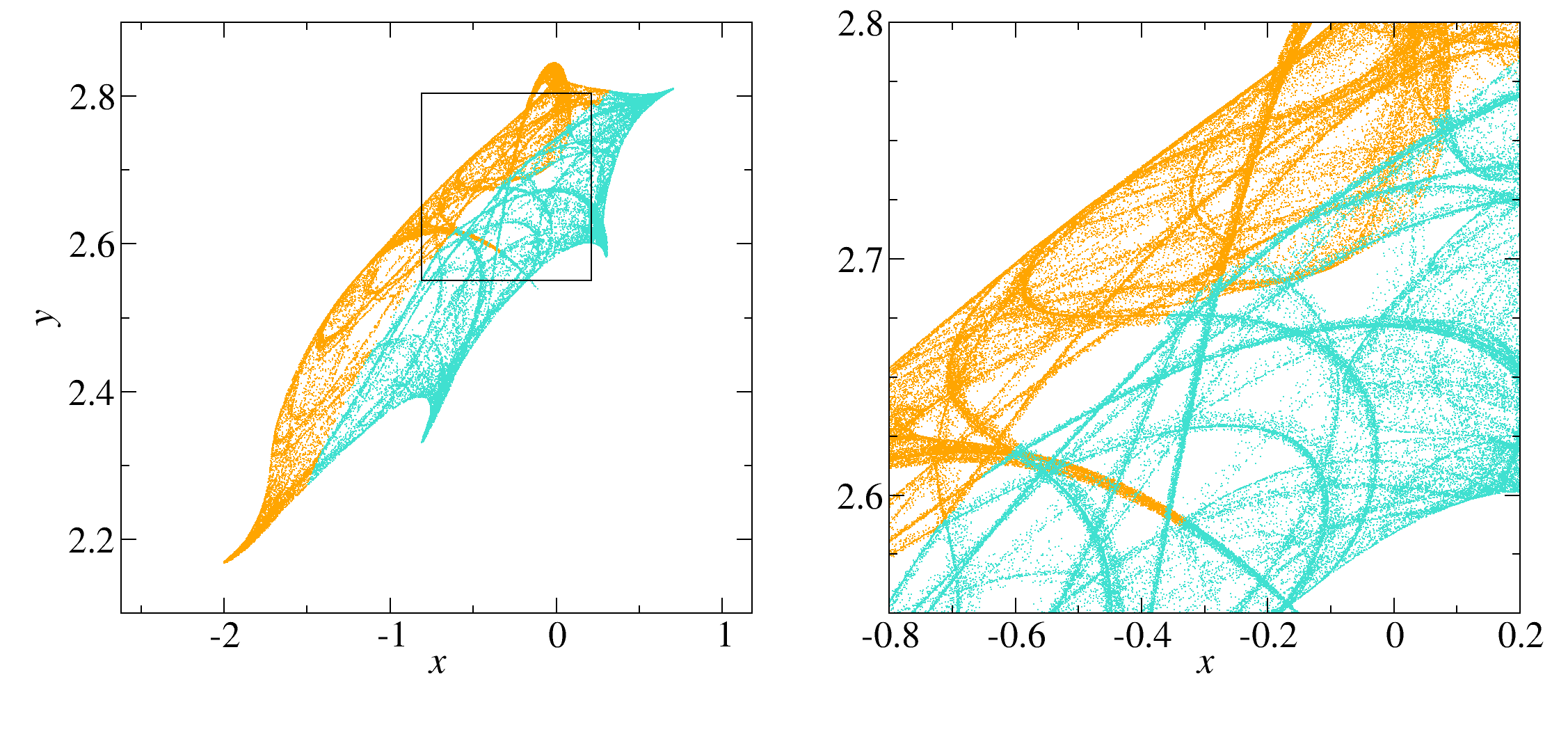} }
\caption{Multivalued velocity fields in 2D maps: projection of the attractor for the dynamics of inertial particles under the Stokes map ($s=0.3$) onto physical space. Colors correspond to particles with $\delta_x >0$ (orange) and $\delta_x <0$ (cyan). The fluid map is the blinking vortex-source system, with parameters $\kappa=-400$, $Q=-20$, $a=1$, and $\tau=0.1$. The right panel is a magnification of the rectangular region in the left panel.}
\label{fig:4}       
\end{figure}

\item{\em Behavior of the Lyapunov dimension of the attractor where particles accumulate.---}
A third mechanism leading to preferential concentration is the dissipative character of the dynamics given by Eq.~(\ref{mapa}). As a result, the formation of strange attractors in phase space is possible. This mechanism is especially prominent for flows with small correlation time, where no centrifuge effect could be present. A widely used measure to quantify clustering in strange attractors is the {\it Lyapunov dimension} $D_L$, also called the Kaplan-Yorke dimension \cite{kaplanyorke}, defined by:
\be
D_L =K+\frac{1}{|h_{K+1}|}\displaystyle\sum_{j=1}^{K}h_{j},
\label{ky}
\ee
where $h_j$ are the Lyapunov exponents displayed in the following ordering: $h_i \geq h_j$ if $i<j$ and $K$ is the largest integer satisfying:
$$
\displaystyle\sum_{j=1}^{K}h_j \geq 0.
$$
For random 2D flows, it has been shown \cite{bec,wilkclustering} that the Lyapunov dimension has the following behavior as a function of $St$. When $St=0$, one observes that $D_L =2$, which is a consequence of the fact that the dynamics then corresponds to that of fluid particles. As $St$ increases, $D_L$ first decreases as a manifestation of preferential concentration. It achieves a minimum for finite $St$ and then starts increasing and asymptotes to the value $4$ as $St \to \infty$ (free-particle limit).

Our Stokes map reproduces qualitatively the described behavior of $D_L$. In the case where the fluid map $\Mf$  corresponds to the area-preserving baker map on the torus (see Appendix A for its expression), $D_L$ can be analytically computed. In fact, the Jacobian matrix is constant in this case, implying that the Lyapunov exponents are simply the natural logarithms of the absolute values of its eigenvalues. The expressions for the Lyapunov exponents are given in Appendix B. Figure \ref{fig:5} shows $D_L$ as a function of $s$ for this choice of $\Mf$.

\begin{figure}
\resizebox{0.99\columnwidth}{!}{%
  \includegraphics{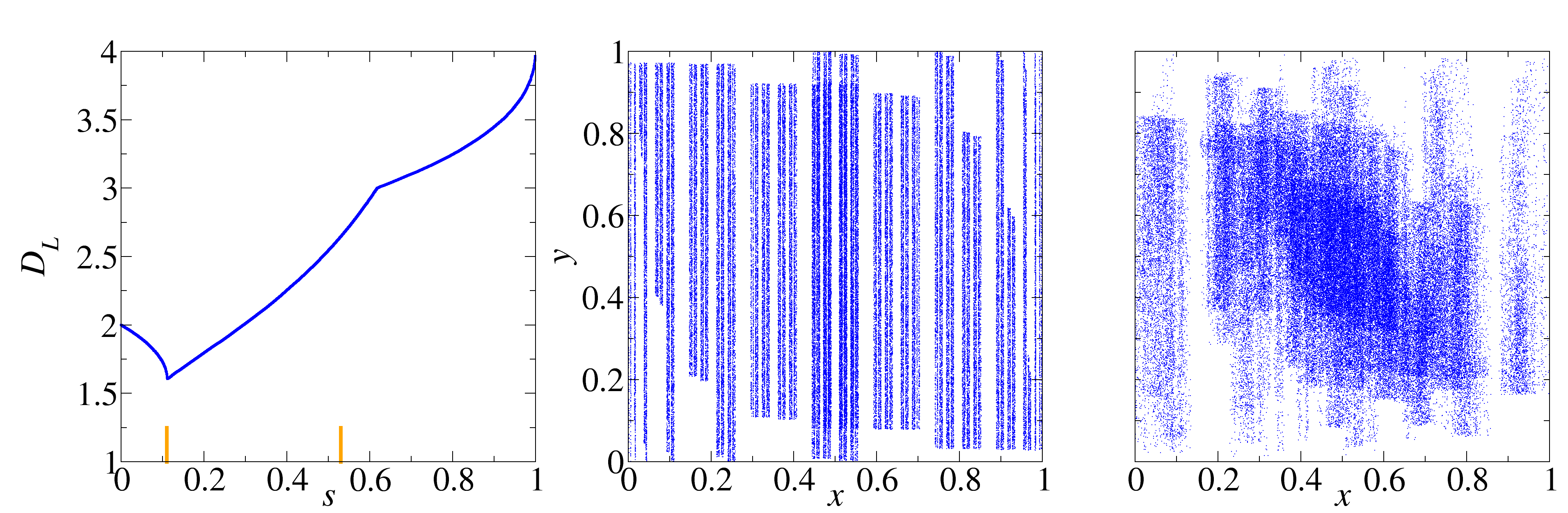} }
\caption{Behavior of the Lyapunov dimension of the attractor where inertial particles accumulate as a function of $s$ (left panel) in the case where the fluid map is the baker map. The two bars perpendicular to the horizontal axis on the left panel correspond to the values of $s$ for which the projection of the attactor onto configuration space is shown: $s=0.11$ (middle panel) and $s=0.53$ (right panel).  }
\label{fig:5}       
\end{figure}

\end{enumerate}

\section{Perspectives}
\label{sec:1}

The Stokes map introduced in this paper could be generalized to account for gravity. Together with the Stokes drag, gravity is a dominant term in the equation of motion for heavy particles in many physical contexts. Other terms such as the added mass can usually be neglected provided that the particle's density is much larger than that of the background fluid\footnote{Note, however, that the condition $R\ll 1$ alone is not sufficient for a negligible history force \cite{daitche2014,ksenia2013,ksenia2016}.}. 

Another topic for future work is the analytical evaluation of all the spectrum of generalized dimensions when the underlying flow is for instance given by a two-scale area-preserving baker map. It is worthwhile noting that, as far as collision rates are concerned, it is the correlation dimension $D_2$ rather than the Lyapunov dimension $D_L =D_1$ which is related to the collision kernel.

The dynamics of collisional growth of inertial particles was studied in the context of chaotic advection by Zahnow {\em et al.} \cite{zahnow1,zahnow2}. The use of the Stokes map could allow the consideration of a much larger number of particles because it reduces the computational cost involved in the time evolution of their trajectories.

In this paper, we have studied the properties of the Stokes map adopting fluid maps which are proxies for chaotically advected flows (the blinking vortex-source and the baker map). We point out, however, that the Stokes map could be also used when the fluid map $\Mf$ corresponds to a turbulent flow. Maps modeling cascades occurring in fully developed turbulent flows were proposed by Hilgers and Beck \cite{hilgers}.

Finally, the formation of attractors for inertial particles in open flows has been reported \cite{vilela,angilella}. It would be interesting to address this phenomenon in the discrete-time modeling.

\acknowledgement{RDV is grateful to Tam\'as T\'el for his encouragement. This paper is dedicated to Professor Ulrike Feudel on the occasion of her 60th birthday.}

\appendix

\section{Expressions for the blinking vortex-source map and the baker map on the torus}

Using the complex coordinate $z$ in the plane of the flow, the blinking vortex-source map reads \cite{telkarolyi}: 
\begin{equation}
z_{n+1}=(z_{n}^{'}-1)\left(1-\frac{\eta}{|z_{n}^{'}-1|^2}\right)^{1/2-i\xi/2}+1, 
\end{equation}
where
$$
z_{n}^{'}=(z_{n}+1)\left(1-\frac{\eta}{|z_{n}+1|^2}\right)^{1/2-i\xi/2}-1.
$$
The parameters $\eta$ and $\xi$ are functions of the source strength $2\pi Q$ (negative for sources and positive for sinks), flow circulation $2\pi \kappa$ (negative for clockwise vortex motion), positions of the sources $\pm a$, and flow period $\tau$: 
$$
\eta=Q\tau/a^2 \mbox{  and  } \xi=\kappa/Q.
$$

The area-preserving baker map on the torus is given by:
\begin{equation}
\Mfb (x_{n}, y_{n})=
(M_{1}(x_{n}, y_{n}), M_{2}(x_{n}, y_{n}))
\equiv\left\{
\begin{array}{ccl}
(x_{n}/2, 2y_{n}), \mbox{     if } 0\leq y_{n}  \leq 1/2,      \cr

(1/2 +x_{n}/2, 2y_{n}-1), \mbox{     if } 1/2 < y_{n}  \leq 1,
\end{array}
\right. 
\label{baker}
\end{equation}
where $x_{n}$ and $y_{n}$ are quantities taken modulo 1.

\section{Eigenvalues of the Jacobian matrix of the Stokes map in the case where $\Mf$ is the baker map on the torus}

In the case where $\Mf=\Mfb$, expression~(\ref{baker}), the Stokes map reads:
$$
\left(
\begin{array}{c}
x_{n+1} \cr
y_{n+1} \cr
\delta_{n+1}^{x} \cr
\delta_{n+1}^{y}
\end{array}
\right)
=
\Msb(x_{n}, y_{n}, \delta_{n}^{x}, \delta_{n}^{y})
\equiv\left(
\begin{array}{c}
[x_{n}+\delta_{n}^{x}] \hspace{1cm} \mbox{ modulo }1 \cr
[y_{n}+\delta_{n}^{y}] \hspace{1cm} \mbox{     modulo }1 \cr
s\delta_{n}^{x} +(1-s)[M_{1}(x_{n+1}, y_{n+1})-x_{n+1}] \cr
s\delta_{n}^{y} +(1-s)[M_{2}(x_{n+1}, y_{n+1})-y_{n+1}]
\end{array}
\right)
$$
The Jacobian matrix is constant and given by:
\be
D\Msb=
\left(
\begin{array}{cccc}
1 & 0 & 1 & 0 \\
0 & 1  & 0 & 1 \\
\frac{s-1}{2} & 0 & \frac{3s-1}{2} & 0 \\
0 & 1-s & 0 & 1
\end{array}
\right)
\ee
The fact that the Jacobian matrix is constant implies that the Lyapunov exponents are simply the natural logarithms of the absolute values of its eigenvalues. The four Lyapunov exponents are given in decreasing order by:

\be
\begin{array}{c}
h_{1}(s) = \ln|\sqrt{1-s}+ 1| \\
h_{2}(s) = \ln|(3s+1+\sqrt{9s^2-10s+1})/4| \\
h_{3}(s) = \ln|(3s+1-\sqrt{9s^2-10s+1})/4| \\
h_{4}(s) = \ln|\sqrt{1-s}- 1|
\end{array}
\ee

With the expressions for the Lyapunov exponents in hands, we can explain the two kinks observed in the graph of $D_L (s)$ (see Fig.~{\ref{fig:5}}, left panel). The first kink occurs at $s=1/9\approx 0.111$, where the $9s^2-10s+1=0$. When this happens, the eigenvalue from which $h_{2}(s)$ is calculated becomes complex and as a consequence $|h_{2}(s)|$ experiences a sharp peak. Since $|h_{2}(s)|$ is then equal to $|h_{K+1}|$ which appears in Eq.~(\ref{ky}) as a denominator, we conclude that $D_L (s)$ must display a sharp minimum (first kink) at $s=1/9$. 

The second kink occurs at $s\approx0.618$, where $D_L =3$ and $K$ (cf. Eq.~(\ref{ky})) jumps from the value 2 to the value 3. At that point, $|h_{K+1}|$  jumps from $|h_{3}(s)|$ to $|h_{4}(s)|$, which is the reason for the kink. We note in passing that no kink is observed at $s\approx 0.296$ where $K$ jumps from the value 1 to the value 2 because there  $|h_{K+1}|$ changes from $|h_{2}(s)|$ to $|h_{3}(s)|$, which are the same (since $h_{2}(s)$ and $h_{3}(s)$ are then complex conjugates).

\vspace{1cm}

\end{document}